\documentclass[aps,prx,reprint,superscriptaddress]{revtex4-1}
\usepackage{graphicx}
\usepackage[export]{adjustbox}
\usepackage[normalem]{ulem}
\usepackage{natbib}
\usepackage{amsmath,array,amssymb}
\usepackage{dcolumn}
\usepackage{bm}
\usepackage{hyperref}
\usepackage{color}
\usepackage{amsmath, amsthm, amssymb, amsfonts}

\graphicspath{{img/}}

\begin{document}
	
\title{Perspectives on quantum transduction}

\author{Nikolai~Lauk}
\affiliation{Division of Physics, Mathematics and Astronomy, California Institute of Technology, 1200 E. California Blvd., Pasadena, CA 91125, USA\\}
\affiliation{Alliance for Quantum Technologies (AQT),  California Institute of Technology, 1200 E. California Blvd., Pasadena, CA 91125, USA\\}

\author{Neil~Sinclair}
\affiliation{Division of Physics, Mathematics and Astronomy, California Institute of Technology, 1200 E. California Blvd., Pasadena, CA 91125, USA\\}
\affiliation{Alliance for Quantum Technologies (AQT),  California Institute of Technology, 1200 E. California Blvd., Pasadena, CA 91125, USA\\}
\affiliation{John A. Paulson School of Engineering and Applied Sciences, Harvard University, 29 Oxford St., Cambridge, MA 02138, USA\\}

\author{Shabir~Barzanjeh}
\affiliation{Institute for Quantum Science and Technology and Department of Physics and Astronomy, University of Calgary, 2500 University Dr. NW, Calgary, AB T2N 1N4, Canada\\}
\affiliation{Institute of Science and Technology Austria, 3400 Klosterneuburg, Austria\\}

\author{Jacob~P.~Covey}
\affiliation{Division of Physics, Mathematics and Astronomy, California Institute of Technology, 1200 E. California Blvd., Pasadena, CA 91125, USA\\}

\author{Mark~Saffman}
\affiliation{Department of Physics, University of Wisconsin-Madison, 1150 University Avenue, Madison, WI 53706, USA\\}

\author{Maria~Spiropulu}
\affiliation{Division of Physics, Mathematics and Astronomy, California Institute of Technology, 1200 E. California Blvd., Pasadena, CA 91125, USA\\}
\affiliation{Alliance for Quantum Technologies (AQT),  California Institute of Technology, 1200 E. California Blvd., Pasadena, CA 91125, USA\\}

\author{Christoph~Simon}
\affiliation{Institute for Quantum Science and Technology and Department of Physics and Astronomy, University of Calgary, 2500 University Dr. NW, Calgary, AB T2N 1N4, Canada\\}

\date{\today}

\begin{abstract}
Quantum transduction, the process of converting quantum signals from one form of energy to another, is an important area of quantum science and technology. The present perspective article reviews quantum transduction between microwave and optical photons, an area that has recently seen a lot of activity and progress because of its relevance for connecting superconducting quantum processors over long distances, among other applications. Our review covers the leading approaches to achieving such transduction, with an emphasis on those based on atomic ensembles, opto-electro-mechanics, and electro-optics. We briefly discuss relevant metrics from the point of view of different applications, as well as challenges for the future.
\end{abstract}
\maketitle	


\section{Introduction}

Transduction refers to the process of converting one form of energy to another. The process of achieving this using individual quantum excitations, referred to as quantum transduction, is an active field of research. Beyond fundamental studies of physics, quantum transduction promises to benefit quantum information science and technology. This is because such transduction allows quantum information to be exchanged between different systems that in general operate at different energy scales and offer their own unique set of attributes. For example, individual infrared photons are excellent carriers of quantum information for fiber optics cables and have been successfully used to transmit quantum information over distances of up to hundreds of kilometers \cite{Gisin2007, Yin2016, Boaron2018, Wengerowsky2019}. On the other hand, many solid state qubit implementations that allow to efficiently perform quantum information processing gates and operations -- such as superconducting circuits, electron spins in quantum dots or NV centers -- typically operate at microwave frequencies. Realizing a quantum transducer that will connect microwave and optical domains, will hence allow to fulfil the DiVincenzo criteria for quantum computing and communication \cite{DiVincenzo_2000}. Such a transducer is particularly important to realize a global quantum internet--a network of quantum computers, or distributed quantum tasks including computing or sensing \cite{Kimble_2008, Simon_2017, Wehner_2018}. Moreover, quantum transduction could be used for efficient detection of microwave photons by exploiting the most efficient detectors for optical photons, or the other way round one could perform non-demolition measurements of optical photons using superconducting qubits coupled to microwave cavities.

If we think about quantum versions of a transducer we most often refer to a faithful transfer of quantum information encoded in one set of bosonic operators $\{ \hat{a}_j \}$ to another set $\{ \hat{b}_j \}$; these could be physically different types of modes, such as photons and phonons, or it could be same types of modes that are disjoint in at least one degree of freedom, such as modes of electromagnetic fields at different frequencies. Quantum frequency conversion between optical fields, first demonstrated in 1992~\cite{Huang_1992}, is now relatively advanced and is applied in many experimental realizations of quantum networks, while microwave-to-microwave photon conversion can be implemented using superconducting circuits \cite{Abdo2013, Lecocq2016}. In this review we want to focus on the particular case of microwave-to-optical quantum transducers. An obvious problem is that the modes have very different frequencies which results in highly off-resonant interactions. One possible way to bridge the five orders-of-magnitude wide energy gap is to use an intermediate system that coherently couples to both microwave and optical modes. Often coupling to such a mediator system results in a non-linear optical interaction. By driving one of the modes with a coherent input the system becomes a parametric oscillator that is described by an effective beam-splitter like Hamiltonian 
\begin{align}
    \hat{H}_{eff} = \hbar \Omega g_{eff} \hat{a}\hat{b}^\dagger + \hbar \Omega^* g_{eff}^* \hat{a}^\dagger\hat{b},\label{Eq:Heff}
\end{align}
where $\Omega$ is proportional to the coherent drive (usually a laser) that provides the required energy and $g_{eff}$ is the effective coupling strength between the optical ($\hat a$) and microwave ($\hat b$) modes. 
There has been a variety of proposals using different kinds of mediating systems; it includes optomechanical systems \cite{, Stannigel_2010, Andrews_2014, Bagci_2014, Barzanjeh_2012, SafaviNaeini_2011, Tian_2010, Wang_2012, Tian_2012, Hill_2012}, atomic ensembles \cite{Imamoglu_2009, Verdu_2009, Hafezi_2012, OBrien_2014, Williamson_2014, Fernandez_2015}, electro-optical systems \cite{Tsang_2010, Soltani_2017}, magnons \cite{Hisatomi_2016}, and others \cite{Das_2017, Elfving_2018, Tsuchimoto_2017}.

In this perspective article we focus on the most widely-researched transduction approaches. 
We review quantum transduction based on atomic ensembles in section II, opto-electro-mechanics in section III, and electro-optics in section IV. In section V we briefly discuss other approaches. Finally in section VI we discuss criteria to assess the performance of quantum transducers from the point of view of different applications, and we end with a discussion of possible future directions and challenges that should be tackled moving forward.



\section{Atomic ensemble based approaches}


The basic idea of ensemble based transducers is to exploit the fact that many atomic systems can have both, microwave and optical, transitions. Usually these transitions are located at different positions in the atomic spectrum and hence a classical optical field has to be used to connect these two transitions in a coherent way. The ability of having optically and microwave addressable transitions is a common feature in atomic systems and hence there is a vast variety of proposed systems to implement a microwave-to-optical transducer ranging from a gas ensemble of neutral atoms, to an ensemble of ions doped into a solid host crystal, to atomic like crystal defects such as NV color centers in diamond. Here we exemplarily discuss some transduction protocols proposed for trapped atomic ensembles and ensembles of rare-earth ions doped into optically transparent crystals. 

\subsection{Ensemble of trapped neutral atoms}

Cold, optically or magnetically trapped neutral atoms offer a pristine system in which transitions in both the microwave and optical regimes can be driven with high fidelity. As such they are a natural setting for the generation of the nonlinearities required for single-photon microwave-to-optical transduction~\cite{Zibrov2002}. In addition the availability of atomic states with long coherence times makes it possible to combine transduction of quantum optical fields with quantum memories that will form the basis for quantum network nodes. 

While electric dipole atomic transitions at optical frequencies are relatively strong, at microwave frequencies the transitions that couple to atomic ground states are of magnetic dipole character, and are much weaker. Consequently, a pronounced challenge with this platform is the ability to engineer a sufficiently high vacuum Rabi frequency of the microwave transition in a physical system that is simultaneously compatible with laser cooling and trapping \textit{and} a cryogenic environment. To meet this challenge either the atom has to be positioned extremely close to the microwave source or resonant cavity, or an ensemble of $N$ atoms should be used to gain a $\sqrt N$ collective enhancement of the coupling. 

The alternative is to excite atoms to Rydberg states which have microwave frequency electric dipole allowed transitions with very large dipole moments, such that strong interactions of single atoms with single microwave photons in free space are possible. This 
characteristic of Rydberg atoms was recognized long ago and exploited for detection of single quanta~\cite{Ducas1979,Figger1980}. Since the atom-microwave coupling via Rydberg states is approximately a factor of $10^6$ larger~\cite{Pritchard2014}, than the coupling via a hyperfine transition in the ground state manifold, a single Rydberg-excited atom can provide the same interaction strength as $10^{12}$ ground state atoms. 

\begin{figure}[t]
\centering
\includegraphics[width=8.6cm]{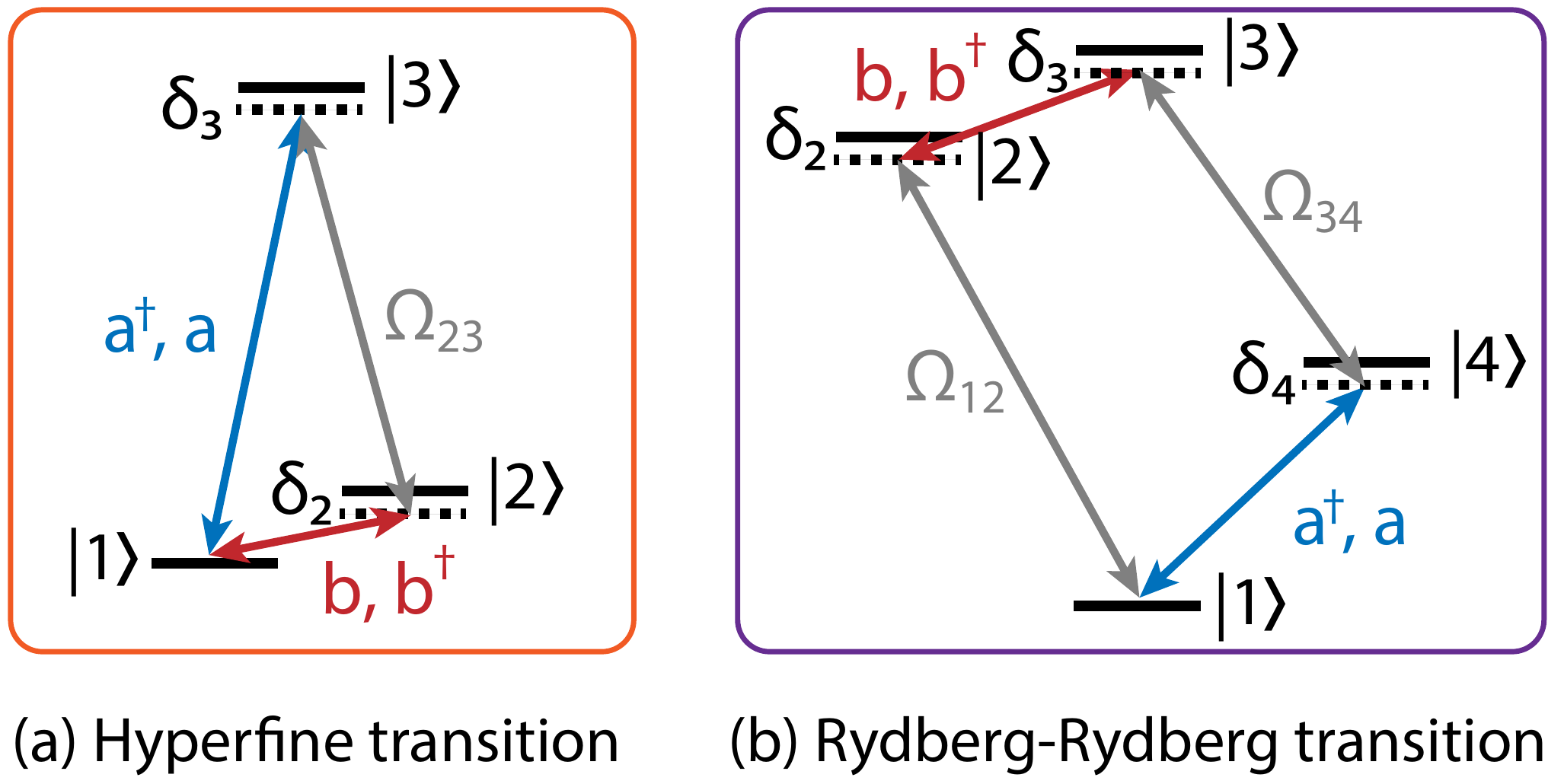}
\caption{\textbf{Level schemes for cold atom-based transduction}. (a) Three-wave mixing with the microwave transition defined by two hyperfine ground states $|1\rangle$ and $|2\rangle$. Both ground states can be optically coupled to an electronic excited state $|3\rangle$. Generically, the fields are detuned from the intermediate states, but maintain a three-photon resonance. (b) One variation of schemes where the microwave transition is defined two highly-excited Rydberg states $|2\rangle$ and $|3\rangle$. Such schemes require four- or higher-wave-mixing. The atomic population primarily resides in $|1\rangle$ in both schemes. Note, however, that while in (a) $|1\rangle$ is coupled to both the optical and microwave field, in (b) it is coupled to only the optical field.}
\label{fig:JCMSFigure}
\end{figure}  

We proceed by highlighting the two primary level schemes employed in transduction experiments with cold, neutral atoms. The first and simplest approach is based on three-wave mixing~\cite{Hafezi_2012}. It employs a microwave transition between a pair of hyperfine ground states in an alkali atom; both of which are optically coupled to the same electronically-excited state (see Fig.~\ref{fig:JCMSFigure}a). The microwave field $\hat b$ ($\hat{b}^\dagger$) annihilates (creates) an excitation of the hyperfine transition $|1\rangle-|2\rangle$, and the optical field $\hat a$ ($\hat{a}^\dagger$) annihilates (creates) an excitation of the optical transition $|1\rangle-|3\rangle$. Generically, there will be a finite detuning between these fields and their corresponding atomic transitions which we call $\delta_{2}$ and $\delta_{3}$, respectively. This system is amenable to pulsed or continuous-wave (CW) operation. Note that the atomic population is predominantly in $|1\rangle$, which is directly attached to both the microwave and optical transitions. 

Crucially, the magnetic dipole moment of the hyperfine transition $|1\rangle-|2\rangle$ is only $\mu\approx1$ $\mu_\text{B}$ (Bohr magneton). Accordingly, even if the atom is $\approx5$ $\mu$m from the surface of a superconducting resonator, the vacuum coupling strength for a single atom would only be $g_\mu/2\pi\approx50$ Hz~\cite{Kim2011,Hoffman2011,Hafezi_2012}. Since superconducting quantum circuits operate with $\approx$MHz single-photon bandwidths~\cite{Devoret2013}, an ensemble of $N\approx10^9$ atoms would be required to obtain a sufficiently large $\sqrt{N}$ collective enhancement~\cite{Verdu_2009}. While such values of $N$ can easily be reached with atomic impurities in solid-state crystals~\cite{Probst_2013,Zhong2017}, this requirement is very daunting for the cold, trapped atom-based platform.

The second and more commonly pursued approach to transduction with cold, trapped atoms relies on microwave transitions between two highly-excited Rydberg states~\cite{Petrosyan2009,Hogan2012,Pritchard2014,Kiffner2016,Gard2017,Han2018,Vogt2019}. (Note that some of these references focus primarily on the microwave-coupling step, which is certainly the most challenging.) Such a transition is employed in a four-wave-mixing scheme (see Fig.~\ref{fig:JCMSFigure}b), or in some cases even six- or seven-wave-mixing. Rydberg-Rydberg transitions have extremely large electric dipole moments that scale as 
$n^2~ e~a_0 $ with $n$ the principal quantum number, $e$ the electronic charge, and $a_0$ the Bohr radius. 
The microwave frequency of the S$-$P transition for a given $n$ scales as $1/n^3$, so the desired frequency can be selected by choosing the principle quantum number $n$ of the Rydberg levels appropriately. For a microwave frequency of $f_\mu\approx5$ ($17$) GHz in cesium (Cs), $n\approx90$ ($60$) would be selected. Typical electric dipole moments of such transitions are $d\gtrsim1000$ $e~a_0$, such that the coupling strength of a single atom at a similar distance of $\approx5$ $\mu$m is $g_\mu/2\pi\approx1$ MHz~\cite{Pritchard2014,Gard2017}. Hence, an ensemble of atoms may not be necessary, and several efforts focus on the use of a single atom. 

However, magneto-optical cooling and optical trapping of atoms -- one or many -- within several $\mu$m of a superconducting waveguide is highly nontrivial. Recall that the superconducting state is easily destroyed upon the absorption of excess photons or magnetic fields~\cite{Devoret2013}. Moreover, the use of Rydberg atoms near surfaces introduces additional challenges associated with their large DC polarizabilities from residual electric fields~\cite{Saffman2010}. While schemes have been devised to reduce the sensitivity of the Rydberg states to these fields~\cite{Booth2018}, alternative approaches in which the atoms can remain far from any surfaces are highly desirable.

Of course as the atom(s) are moved further from a surface, the coupling strength of the microwave transition decreases rapidly. Naively, one might assume that this decrease in the coupling strength can be compensated by a collective enhancement via the use of an atomic ensemble. However, here in lies a subtle yet crucial point. In contrast to three-wave mixing schemes considered above, in the Rydberg-atom-based scheme operating in the single-photon regime there is no steady-state atomic polarization in either of the states coupled to the microwave field (i.e. $|2\rangle$ and $|3\rangle$). This is precluded by the relatively short lifetimes, scaling approximately as $n^3$ ($\approx1$ ms at $n=90$ in a 4K cryostat~\cite{Beterov2009}). Thus, there is no collective enhancement of the microwave transition in the single-photon regime. Reference~\cite{Covey2019b} provides a detailed analysis of this effect, and proposes a compromise between coupling strength and distance from surfaces. Note, however, that a collective enhancement could be engineered in a resonant, pulsed regime~\cite{Petrosyan2019}. Even so, it is difficult to continuously excite a dense sample of Rydberg atoms in the multi-photon regime due to the blockade effect~\cite{Saffman2010}.

It is desirable that the wavelength of the optical photon lie in the telecommunications window ($\sim1.25$ $\mu$m to $\sim1.65$ $\mu$m) for more efficient photon transfer over long fiber links. Generally, most atom and atom-like emitters have strong optical transitions in the visible band where the wavelengths are much shorter, and frequency conversion into the telecom window is often required. Erbium ions are a notable exception to this trend with transitions at $1.54$ $\mu$m; however, these transitions are very weak and must be Purcell- or collectively-enhanced with an optical resonator to achieved desired bandwidths compatible with superconducting circuits. In alkali atoms telecom-band transitions are only available between high-lying states, so complex schemes involving six or seven internal levels are required to make use of them for transduction~\cite{Pritchard2014,Kiffner2016,Gard2017,Han2018,Vogt2019}. Recently, a CW four-wave-mixing scheme based on an ensemble of alkaline-earth(-like) ytterbium (Yb) atoms was proposed~\cite{Covey2019b} in which a strong transition in the telecommunication E-band at 1389 nm is employed~\cite{Covey2019}.


Experimental progress on cold atom-based transduction lags substantially behind leading approaches in this field, primarily because of its relative complexity. Early work in this field focused on atomic beams~\cite{Hogan2012,Thiele2014,Thiele2015,Stammeier2017} rather than cold, trapped atoms; however, a number of efforts based on the latter are now in progress. References~\cite{Kim2011,Hoffman2011,Hafezi_2012,Grover2015,Hoffman2015,Hattermann2017} provide an overview of approaches based on three-wave mixing using a ground-state hyperfine transition. References~\cite{Petrosyan2009,Hogan2012,Pritchard2014,Kiffner2016,Gard2017,Han2018,Vogt2019} provide an overview of ideas and recent experimental efforts with Rydberg states, but we emphasize that this list is not exhaustive. 

The first demonstration of the microwave-to-optical conversion with Rydberg atoms to our knowledge was performed in 2018 in Ref.~\cite{Han2018}. However, the efficiency was low ($\eta\sim0.003$) and the conversion was performed in the classical regime with many extra photons. Nevertheless, a respectable conversion bandwidth of $\bar{\Gamma}\approx4$ MHz was observed. In an improvement upon this first result, the same group demonstrated a higher efficiency of $\eta\sim0.05$~\cite{Vogt2019}, albeit still in the classical regime with many excess photons. Using numerical simulations of their system the authors conclude that the conversion efficiency could be increased up to $\eta \approx 0.7$ with corresponding conversion bandwidth of $\bar{\Gamma}\approx15$ MHz. While there have been few demonstrations of microwave to optical transduction with Rydberg atoms to date, more effort has focused on the related problem of coupling atoms to microwave fields - specifically superconducting resonators. Reference~\cite{Verdu_2009}, for instance, demonstrated the strong coupling of an ultracold gas to a superconducting waveguide cavity using a hyperfine transition in the ground state (not Rydberg). They observed $g_\text{eff}/2\pi\sim40$ kHz, which is large compared to the cavity linewidth $\kappa/2\pi\sim7$ kHz.

\subsection{Rare-earth-ion-doped crystals}

Ensembles of rare earth-ions (REI) doped into transparent crystals are an appealing platform to devise transducer interconnects. They are well known for their long optical coherence times and form the basis for solid state implementations of optical quantum memories. More recently their spin coherence times were measured and analyzed in more detail, and coherent ensemble coupling to microwave cavities was demonstrated \cite{Probst_2013} allowing the development of transducer proposals in this medium. First transducer proposals for REI ensembles were made a few years ago \cite{OBrien_2014, Williamson_2014}. Both proposal suggest to use $\mathrm{Er}^{3+}$ ions doped into yttrium orthosilicate (YSO) crystal due to its prominent optical transition in the telecom wavelength region at 1536 nm. With half-integer total spin $\mathrm{Er}^{3+}$ belongs to the so called Kramers ions and as such has a doubly degenerate ground and optical excited state. This degeneracy could be lifted by applying an external static magnetic field allowing for a magnetic dipole transition in the microwave range. Another advantage of $\mathrm{Er}^{3+}$ is its relatively high magnetic dipole moment, up to 15 $\mu_B$ in YSO host crystal~\cite{Sun_2008}. However, magnetic dipole transitions are in general considerably weaker than electric dipole transitions and a high Q microwave resonator as well as large number ensembles have to be used to enhance the coupling to microwave fields. 

\begin{figure}[t]
 	\includegraphics[width=\columnwidth]{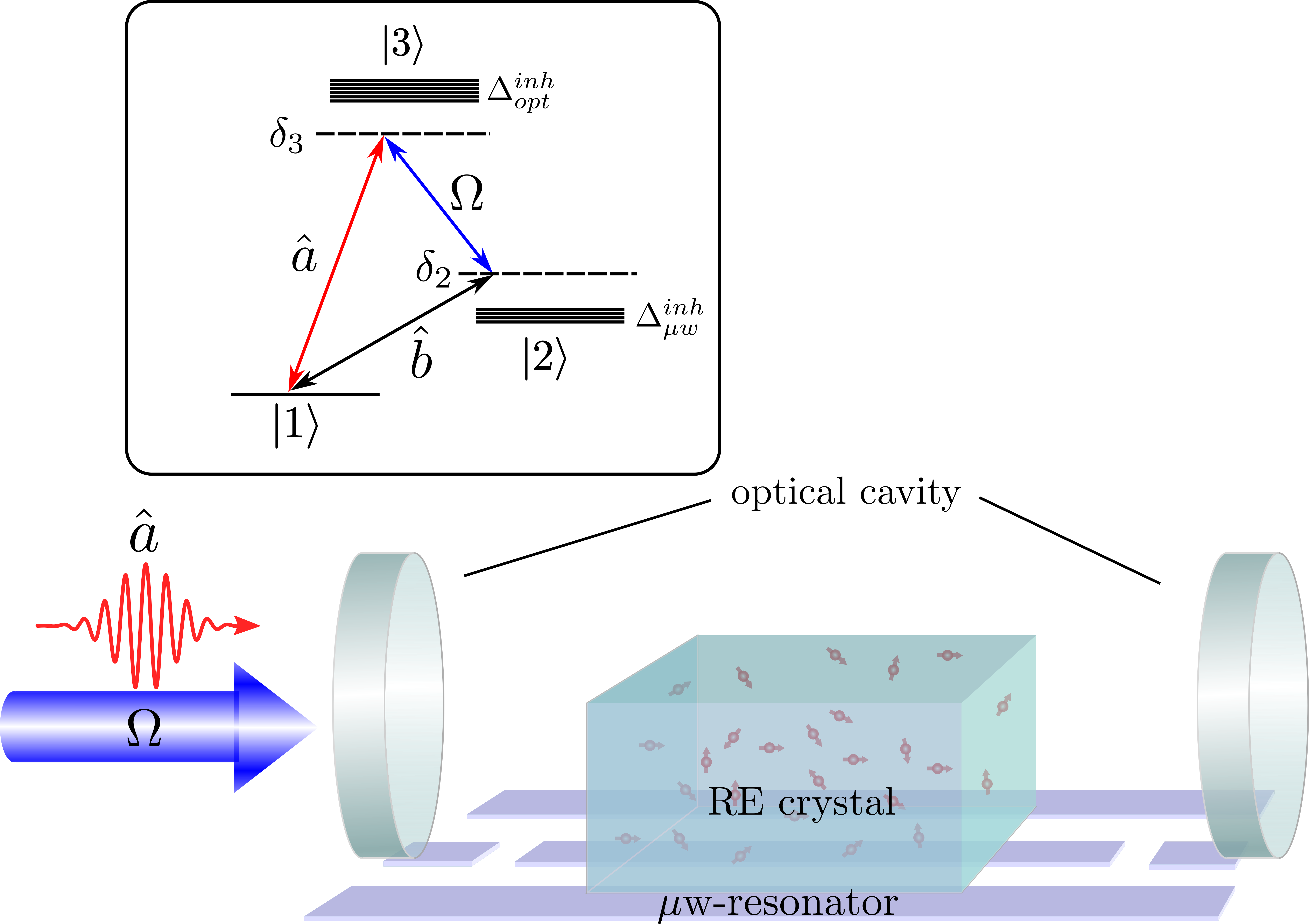}
 	\caption{Schematic setup for a rare-earth crystal based transduction. An optically transparent crystal doped with rare-earth ions is placed within an optical cavity and a microwave resonator. Each ion can be modelled as $\Lambda$-system consisting of three levels, see inset. The state $|3\rangle$ couples via electric dipole transition to the states $|1\rangle$ and $|2\rangle$, which in turn are coupled via magnetic dipole transition with each other. Due to different crystal environments at the ion's locations the transitions are inhomogeneously broadened, as indicated in the inset by $\Delta^{inh}_{opt}$ and $\Delta^{inh}_{\mu w}$.}
 	\label{Fig:REsetup}
\end{figure}

The system can be considered as an ensemble of three level atoms in $\Lambda$-type configuration, with two lower lying spin levels and one common optically excited state, similar to the three-wave mixing scheme for the trapped atoms (cf. Fig. \ref{Fig:REsetup}). But unlike the trapped atom ensembles, the transitions in REI ensembles have large inhomogeneous broadening due to slightly different local environment in the host crystal. One way to mitigate the detrimental effects of large inhomogeneous broadening is to use optical and microwave cavities that are far detuned from the resonant transitions (as suggested in \cite{Williamson_2014}), with detunings $\delta_3$ and $\delta_2$ being larger than the inhomogeneously broadened linewidths of corresponding transitions. In this case the matter part can be adiabatically eliminated from the equations of motion and one is left with an effective nonlinear interaction between the classical optical field, quantum optical cavity field and quantum microwave cavity field with the interaction Hamiltonian \eqref{Eq:Heff}, 
where $\Omega$ is the Rabi frequency of a classical field driving the $|2\rangle - |3\rangle$ transition, and $\hat a$ ($\hat a^\dagger$) and $\hat b$ ($\hat b^\dagger$) are the annihilation (creation) bosonic operators for the optical and microwave cavity respectively. The effective coupling strength $g_{eff}$ depends on the collective coupling strength of the optical and microwave transitions and is inversely proportional to the detunings. The fact that the detunings should be large results in the usually weak effective coupling strength $g_{eff}$ and efficient conversion requires large cooperativity factors for both cavities.

The scheme above, with classical field being constantly present, is best suited for a conversion of CW fields. But pulsed operations could be implemented in the same medium as well, as was shown in \cite{OBrien_2014}. In this proposal the conversion of an optical photon to a microwave photon is achieved by first mapping an incoming single photon pulse onto an optical matter excitation and subsequently transferring this excitation to a spin excitation by means of a series of classical laser pulses. Afterwards the spin excitation is resonantly coupled to a microwave cavity which leads to a coherent mapping of the spin excitation to a microwave photon. To store the incoming photon into a collective matter excitation the authors propose to use gradient echo quantum memory protocol, a type of a controlled reversible inhomogeneous broadening (CRIB) quantum memory protocol \cite{CRIB-memory}, which together with atomic frequency comb (AFC) protocol \cite{AFC-memory} belongs to echo type quantum memories and was especially designed for systems with long optical coherence times and large inhomogeneous broadening, such as rare-earth doped crystals. The idea of the gradient echo memory protocol is to induce controllable inhomogeneous broadening in the system by applying a field gradient. Large inhomogeneous broadening leads to a fast dephasing (much faster than the excited state lifetime) of the collective excitation that couples to the electromagnetic field resulting in the effective storage of the photon. In order to transport this optical excitation to a rephased spin excitation a series of $\pi-$pluses is applied between the spin and excited levels. The timing of the pulses is such that at the end of the rephasing procedure the system is left in the collective spin state that interacts resonantly with the microwave cavity. The overall efficiency of the protocol is given by $\eta=\eta_S\eta_T$, where $\eta_S$ is the storage efficiency of the optical photon into the symmetric spin excitation and $\eta_T$ describes the transport efficiency of this spin excitation to a microwave photon inside the cavity. The storage efficiency is bounded by the spatial overlap of the spin excitation with the microwave cavity mode. Using gradient echo scheme where the field gradient is applied along the propagation direction of the optical photon this overlap can be maximized by spectral tailoring of the incoming photon, since in this configuration the spatial profile of the spin excitation is given by the frequency spectrum of the optical photon.


A more recent proposal \cite{Welinski_2018} suggest modification of the proposal by O'Brien et al. \cite{OBrien_2014} by using the Zeeman levels of the optically excited state instead of the ground state. The modest ratio between the coupling strength and the decoherence rate limited the conversion efficiency in the original proposal \cite{OBrien_2014} to $\sim90\%$. Using the sublevels in the optically excited state manifold have advantage of longer spin coherence times due to reduced spin-spin interaction with the neighboring ions and could potentially improve the overall conversion efficiency near unity.


Most experimental investigations for light-matter coupling using rare-earth crystals have focused on demonstrations of quantum memories for light or microwaves. There are several reviews of quantum memories for light \cite{Tittel_2010, Lvovsky_2009}, and many works on quantum-oriented microwave or radio-frequency coupling using bulk and stripline resonators \cite{Probst_2013}. Nevertheless, there has been little experimental work using enesmbles of rare earth ions for microwave to optical transduction.

The first efforts came from the Longdell group of Otago that followed a route based on off-resonant fields and three-wave mixing \cite{Williamson_2014, Fernandez_2015}. In one of their experiments a cylindrical sample of erbium
doped YSO crystal sat inside a shielded loop gap resonator and optical Fabry-Perot resonator in a superconducting magnet at 4.6 K under 146 mT field. Using this system it was possible to demonstrate microwave to optical telecom-wavelength conversion for classical input fields with an efficiency of order $10^{-5}$ \cite{Fernandez_2017}. The optical cavity in the system has provided an enhancement in the conversion efficiency of nearly $10^4$ compared to its counterpart without an optical cavity \cite{Fernandez_2015}. The authors predict that by matching the impedance of the microwave and optical cavities as well as by lowering the temperature to mK one should obtain near unit conversion efficiency.

Another experimental effort comes from the Faraon group at Caltech that has pioneered approaches using nanophotonic waveguide and cavities used focused ion beam milling of yttrium vanadate crystals \cite{Bartholomew2019}. For transduction, their approach involves using Yb ions under weak magnetic field. Yb ions have a non zero nuclear spin, which results in a simple hyperfine strucutre (characterized by a nuclear spin of 1/2) that also features a long coherence lifetime due to zero first order interaction of the spin with its surrounding magnetic field bath. In their design the nanophotonic components are positioned within microwave coplanar waveguides and cavities that allow the necessary microwave coupling. Optically detected microwave resonance spectroscopy has illustrated the necessary coupling between light and microwaves.


\section{Opto-electro-mechanics}


The most well-known and accomplished advancements to optical-to-microwave transduction involve the simultaneous coupling of light to mechanical motion, i.e. optomechanics, and the coupling of their motion to microwaves, i.e. electromechanics. Specifically, this can be achieved by the interaction of microwaves in an LC circuit by using electrostatic forces, e.g. capacitance, which displaces the boundaries of a nanomechanical resonator and using light in a cavity that is coupled through photoelasticity and the accompanying displacement of its boundaries. These couplings can be inferred through variations in the resonances of both the optical and microwave cavities by the mechanical motion. The small size of the mechanical objects reflects the desire to obtain strongly coupled systems as well as the requirement for low mechanical stiffness, allowing large displacements. Nonetheless, the masses of current systems, which is limited by the device sizes of order 100 nanometers, restrict transduction bandwidths compared to other approaches.

\subsection{Theoretical model of the mechanical based photon conversion}\label{seq:OptomechanicalTheory}

\begin{figure}[ht]
 	\includegraphics[width=\columnwidth]{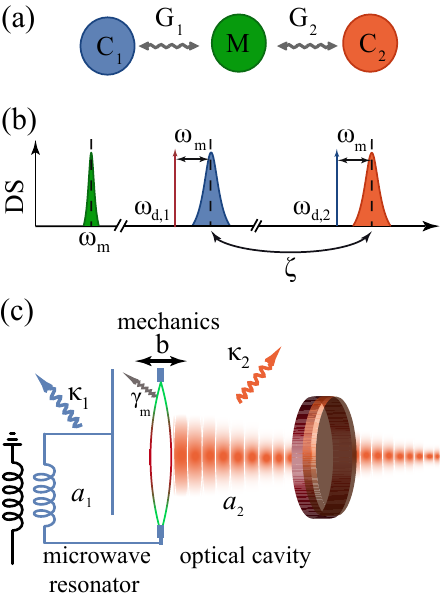}
 	\caption{(a) Modes coupling diagram describing the coupling of the microwave mode $C_1$ and the optical mode  $C_2$ to the mechanical resonator mode $M$. The coupling strength between microwave (optical) field and the mechanical resonator is $G_1$ ($G_2$). (b) Density of states (DS) of the mechanical resonator $\omega_m$, microwave resonator $\omega_{c,1}$, optical cavity $\omega_{c,2}$, and drives $\omega_{d,i}$. The electro-optomechanical interaction meditates the microwave-to-optical transduction with conversion efficiency $\zeta$. (c)  Schematic representation of the microwave-to-optical photon conversion using a mechanical resonator. The mechanical vibration of a movable membrane, with annihilation operator $\hat b$, is shared between a microwave resonator, with annihilation operator $\hat a_1$, and an optical cavity with annihilation operator $\hat a_2$. The membrane acts as one of the mirrors of the optical cavity while being capacitively coupled to a superconducting microwave resonator. Here, $\gamma_m$, $\kappa_1$ , and $\kappa_2$ are the damping rates of the mechanical resonator, the microwave resonator, and the optical cavity, respectively.  The vibration of the membrane modulates the resonance frequencies of the optical cavity and microwave resonator.}
 	\label{FigOMC}
\end{figure}

Before presenting the recent advances of the optomechanical based photon conversion we theoretically explain how the mechanical resonator facilitates the photon conversion between microwave and optical domains. Fig. \ref{FigOMC}(a) shows the modes coupling diagram of a double-cavity optomechanical system in which a microwave resonator mode $C_1$ with resonance frequency $\omega_{c,1}$ and an optical cavity mode $C_2$ with resonance frequency $\omega_{c,2}$, are simultaneously coupled to the vibrational mode of a mechanical resonator $M$ with frequency $\omega_m$.  In Fig. \ref{FigOMC}(c) we schematically show a circuit representing this mode coupling in which a mechanical resonator forms  one of the mirrors of the optical cavity while it capacitively coupled to a superconducting microwave resonator. The Hamiltonian describing this tripartite interaction is given by \cite{Aspelmeyer2014}
\begin{eqnarray}
\hat{H} &=&\hbar\, \omega _{m}\hat{b}^{\dagger }\hat{b}+\hbar \sum_{j=1,2}\Big[\omega _{c,j}\hat{a}_{j}^{\dagger }\hat{a}_{j}+ g_{0,j}(\hat{b}^{\dagger }+\hat{b})\hat{a}%
_{j}^{\dagger }\hat{a}_{j}\nonumber\\
&+&\mathrm{i}\, \mathcal{E}_{j}(\hat{a}_{j}^{\dagger }e^{-\mathrm{i}%
\omega _{\mathrm{d},j}t}-\hat{a}_{j}e^{\mathrm{i}\omega _{\mathrm{d},j}t}) \Big].
\end{eqnarray}
where $\hat{b}$ is the annihilation operator of the mechanical resonator, $\hat{a}_{j}$ is the annihilation operator for
resonator $j$ whose coupling rate to the mechanical resonator is $g_{0,j}$. As shown in Fig. \ref{FigOMC}(b) the optical cavity and microwave resonator are driven by external coherent pumps with amplitude
$\mathcal{E}_{j}$ and frequencies $\omega _{\mathrm{d},j}$~\cite{Barzanjeh2011a}. 

Moving in the interaction picture with respect to $\hbar \omega _{\mathrm{d,1}}\hat a_{%
\mathrm{1}}^{\dagger }\hat a_{\mathrm{1}}+\hbar \omega _{\mathrm{d}%
,2}\hat a_{2}^{\dagger }\hat a_{2}$, neglecting terms oscillating at $\pm 2\omega
_{\mathrm{d,j}}$, and linearized the Hamiltonian  by expanding the resonator modes
around their steady-state field amplitudes, $\hat{c}_{j}=\hat{a}_{j}-\sqrt{%
n_{j}}$, where $n_{j}\gg 1$ is
the mean number of intracavity photons induced by the cavity
pumps~\cite{Barzanjeh2011a,Barzanjeh2017}, result in the linearized system Hamiltonian 
\begin{equation}
\hat{H}=\hbar \sum_{j=1,2}G_{j}(\hat{b}\mathrm{e}^{-\mathrm{i}%
\omega _{m}t}+\hat{b}^{\dagger }\mathrm{e}^{\mathrm{i}\omega _{m}t})(\hat{c}%
_{j}^{\dagger }\mathrm{e}^{\mathrm{i}\Delta _{j}t}+\hat{c}_{j}\mathrm{e}^{-%
\mathrm{i}\Delta _{j}t}),  \label{ham3}
\end{equation}
where $\Delta
_{j}=\omega _{c,j}-\omega _{\mathrm{d},j}$ is the cavity/resonator-pump detuning and $G_j=g_{0,j}\sqrt{n_j}$ is the multiphoton optomechanical cavity rate.  By selecting the detuning parameter we can either choose the beam splitter or parametric like interaction in the optomechanical system. By setting the effective resonator detunings
so that $\Delta _{1}=-\Delta _{2}=-\omega _{m}$ and neglecting the
terms rotating at $\pm 2\omega _{m}$, the Hamiltonian (\ref{ham3}) reduces to
\begin{equation}
\hat{H}_\mathrm{TMS}=\hbar G_{1}(\hat{c}_{1}\hat{b}+\hat{b}^{\dagger }\hat{c}%
_{1}^{\dagger })+\hbar G_{2}(\hat{c}_{2}\hat{b}^{\dagger }+\hat{b}\hat{c}_{2}^{\dagger }),  \label{hampara}
\end{equation}
The first two terms of the above Hamiltonian are responsible for generating entanglement between photonic excitation of the microwave mode $C_1$ and mechanical mode $M$ while the last two terms represent a beam splitter like interaction which exchanges the excitation between the optical mode $C_2$ and mechanical mode $M$. This specific form of the interaction can be used to generate entanglement or two mode squeezing (TMS) between the output radiation of the cavities \cite{Barzanjeh2019, Barzanjeh2015, Barzanjeh2013}. This type of interaction can be used for high-fidelity quantum states transfer between optical and microwave fields in forms of continuous variable quantum teleportation \cite{Barzanjeh_2012}.

On the other hand, by selecting the effective resonator detunings $\Delta _{1}=\Delta _{2}=\omega _{m}$ and neglecting the
terms rotating at $\pm 2\omega _{m}$, the Hamiltonian (\ref{ham3}) reduces to \cite{Tian_2010}
\begin{equation}
\hat{H}_\mathrm{BS}=\hbar G_{1}(\hat{c}_{1}\hat{b}^{\dagger }+%
\hat{b}\hat{c}_{1}^{\dagger })+\hbar G_{2}(\hat{c}_{2}\hat{b}^{\dagger }+%
\hat{b}\hat{c}_{2}^{\dagger }),  \label{hambeam}
\end{equation}%
indicates beam splitter-like interaction between mechanical degree of freedom
and microwave resonator (optical cavity) mode, appropriates for the photon conversion between microwave resonator and optical cavity. In the photon transduction process, the microwave photons indicated by $\hat a_{1}$ are down-converted into the mechanical mode at frequency, i.e.~$\hat a_1\xrightarrow{\hat H \propto \hat a_1\hat b^{\dagger}} \hat b$. 
Next, during an up-conversion process the mechanical mode transfers its energy to the optical cavity mode $\hat a_{2}$, i.e.~$\hat b\xrightarrow{\hat H \propto \hat b \hat a^{\dagger}_2} \hat a_2$. 
This process is bidirectional which means the photons of the optical mode can be converted to the microwave mode by reversing the conversion process. 

The photon conversion efficiency between the outputs of the microwave resonator and the optical cavity in the steady state and in the weak coupling regime is given by \cite{Tian_2010, Lecocq2016}
\begin{equation}\label{efficenyeq}
\zeta=\frac{4 C_1C_2}{(1+C_1+C_2)^2},
\end{equation}
where $C_{i}=\frac{4G_i^2}{\kappa_{i}\gamma_m}$ is the optomechanical cooperativity for cavity $i=1,2$ in which $\kappa_i$ is the total damping rate of the microwave and optical cavities, and $\gamma_m$ is the damping rate of the mechanical resonator.  Note that in Eq. (\ref{efficenyeq}) we ignore the internal losses of the optical cavity and microwave resonator. In the limit of equal and large cooperativity $C_1=C_2$ and $C_i/n_m\gg 1$, the unity \textit{coherent} photon conversion can be achieved $\zeta =1$ where $n_m$ is the thermal occupation of the mechanical mode.  The
bandwidth of the conversion is set by the total mechanical damping
$\Gamma=\gamma_m(1+C_1+C_2)$ which is the total back-action-damped linewidth of the mechanical resonator.

\subsection{Photon conversion using mechanics}

Among the initial experiments, Bagci et al. \cite{Bagci_2014} demonstrated a strongly-coupled opto-electro-mechanical transducer using an electrostatic nanomembrane that is displaced out-of-plane using a radio-frequency resonance circuit and is simultaneously coupled to light reflected off its surface. The mechanical resonator used in this experiment was a 500-$\mu$m-square SiN membrane coated with Al. The radio-frequency signals are detected as an optical phase shift with quantum-limited sensitivity. Thermal noise fluctuation and the quantum noise of the light are the two major sources of the noise which both dominated the Johnson noise of the input.

Andrews et al. \cite{Andrews_2014} have shown bidirectional transduction with 10$\%$ efficiency overall albeit with about 1700 noise quanta at cryogenic temperatures. Their system employs a thin SiN membrane that acts as one of the mirrors of an optical Fabry-Perot cavity while being capacitively coupled to a superconducting microwave resonator. The vibration of the membrane simultaneously modulates both optical cavity and microwave resonator which consequently transfers the excitation between microwave and optical modes. This measurement has been done in near resolved-sideband regime in which the mechanical frequency $\omega_m$ exceeds the damping rates of the microwave resonator $\kappa_1$ and the optical cavity $\kappa_2$ set by $4\,\omega_m>\{\kappa_1,\kappa_2\}$. The conversion efficiency in this experiment was limited by the loss of the microwave resonator and the imperfect optical mode matching. The thermal vibrational noise at 4 K temperature as well as the spurious mechanical modes in the membrane were the two main sources of the noise in this experiment. These issues have been resolved in the recent experiment from the same group \cite{Higginbotham2018}. Improving the sample design to remove the unwanted mechanical modes, having low-loss optical cavity and microwave resonator, larger opto and electromechanical coupling rates, and operating the sample below 40 mK temperature result in considerable improvement of the photon conversion to $48\,\%$ with only 38 added noise quanta. 

\subsection{Photon conversion using piezoelectric effect}

Another avenue involves piezoelectricity to achieve the electro-mechanical coupling, which does not involve defining an electro-mechanical capacitor, but instead using the traveling phonons.

Bochmann et al. \cite{Bochmann_2013}, demonstrated bidirectional microwave-to-optical conversion of strong fields using a piezoelectric aluminum nitride optomechanical photonic crystal cavity. The piezeoelectric coupling allowed mechanical strain to couple to microwave fields via an interdigital transducer while a one dimensional optomechanical cavity hosted high quality mechanical and optical modes. In their experiment, coherent mechanical modes are driven through the interdigital transducer and optical read out is provided by an evanescently coupled waveguide. Internal conversion efficiencies are only at the few percent level. 

Another experimental effort using piezoelectrics optomechanics \cite{Balram} comes from the Srinivasan group at NIST that coherently coupled radio frequency, optical, and acoustic waves in an integrated chip. In this experiment an optomechanical cavity with photonics wavelength $1550$ nm and localized phononics mode $2.5$ GHz are placed between two inter-digitated transducers (IDTs). The strong optomechanical coupling rate in the order of $1$ MHz allowed efficient coupling between the optical mode and the localized mechanical breathing mode. The RF excitation is first converted to surface acoustic wave using the IDTs and then routed to the optomechanical cavity using phononic crystal waveguides which ultimately excited the mechanical mode of the optomechanical cavity and therefore facilitated the energy conversion between the optical and radio frequency modes.

In a similar effort Forsch et al. \cite{Forsch2018} have implemented the quantum groundstate microwave-to-optical photon conversion. In this device a one-dimensional optomechanical crystal is coupled to an IDT. The optomechanical cavity supports a breathing mechachanical mode at 2.7 GHz while its photonics mode is in the telecome band. The piezoelectric effect, which creates the electromechanical coupling, links the microwave excitation to the optomechanical cavity, allows quantum noise limited bidirectional conversion with efficiency in the order of $5.5\times 10^{-12}$. 

Similarly, the  Safavi-Naeini's group have demonstrated a low-noise on-chip lithium niobate piezo-optomechanical transducers using acousto-optic modulation  \cite{Jiang2019}. This system provides bidirectional conversion efficiency of $10^{-5}$ with red-detuned optical pump and $5.5\%$ with blue-detuned pump.


\section{Electro-optics}

While opto-electro-mechanical and atomic ensemble-based transduction between light and superconducting qubits has attracted a lot of attention recently, only little interest was shown for the coherent coupling of light and microwaves at the quantum level through electro-optic interactions \cite{Tsang_2010, tsang2011} despite the interaction being widely-used for classical opto-electronic applications \cite{kaminow2013introduction}. This is partly due to the lack of an electro-optic effect in most (centrosymmetric) materials, and a weak electro-optic single-photon coupling strengths offered by bulk (non-centrosymmetric, i.e. $\chi^{(2)}$) non-linear optical systems. However, recent improvements of the quality of optical resonators fabricated from thin, low-loss, non-linear materials, see e.g. Ref. \cite{zhang2017}, or the development of nanoscale evanescently-coupled cavities on bulk crystals, such as in Ref. \cite{witmer2017}, in conjunction with the possibility of large mode overlap between optical microwave resonator fields has established this approach for quantum applications. 

\begin{figure}[t]
    \includegraphics{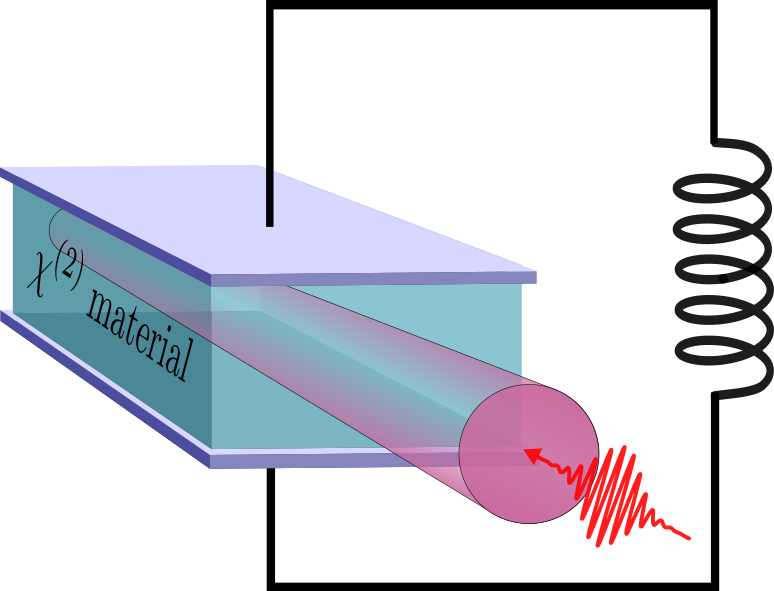}
    \caption{Schematic representation for an electro-optic-based transducer. A $\chi^{(2)}$ nonlinear material is placed inside a microwave resonator and driven with strong light. The microwave photons can be reversibly converted to optical sideband photons.}
    \label{Fig:EOscheme}
\end{figure}

Consider an electro-optic material, i.e. one that mediates energy transfer between electric and optical fields by $\chi^{(2)}$, inside of a microwave and optical resonator. A cartoon schematic representation is shown in Fig. \ref{Fig:EOscheme}. Following the approach of Tsang \cite{Tsang_2010}, the interaction Hamiltonian for the electro-optic effect is given by
\begin{align}
H_i &= -\frac{\hbar}{\tau} \phi \hat a^\dagger \hat a,
\end{align}
where $\hat a$ and $\hat a^\dagger$ are the annihilation and creation operators for the optical cavity mode, respectively, $\tau$ is the optical round-trip time of the optical cavity, and $\phi$ is the single-round-trip phase shift. The round-trip electro-optic phase shift, on the other hand, is given by
\begin{align}
\phi &= \frac{\omega_a n^3 r l}{c d} V,
\end{align}
where $n$ is the optical refractive index inside the electro-optic medium, $r$ is the electro-optic coefficient in units of m$/$V, $l$ is the length of the medium along the optical axis, $d$ is the thickness, and $V$ is the voltage across the medium. For our application, the electro-optic material can be modeled as a capacitor forming part of the microwave resonator so that the voltage can be defined as
\begin{align}
\hat V &= \left(\frac{\hbar\omega_b}{2C}\right)^{1/2}\left(\hat b + \hat b^\dagger\right),
\end{align}
where $\hat b$ and $\hat b^\dagger$ are the microwave annihilation and creation
operators, respectively, $\omega_b$ is the microwave resonant frequency, and $C$ is the capacitance of the microwave resonator. The full Hamiltonian for determining cavity electro-optical dynamics in the (single-photon) strong-coupling regime can be written as \cite{Tsang_2010}
\begin{equation}
\label{eq:Hamiltonian}
\hat{H} = 
\hbar \omega_{a} \hat{a}^{\dagger}\hat{a} 
+ \hbar \omega_{b} \hat{b}^{\dagger}\hat{b} 
- \hbar g_0 \left( \hat{b} + \hat{b}^{\dagger} \right) 
\hat{a}^{\dagger}\hat{a}.
\end{equation}

The electro-optic coupling coefficient is given by
\begin{equation}
g_0 = \tfrac{\omega_{a} n^{3} r l}{c \tau d} \sqrt{\tfrac{\hbar \omega_{b}}{2 C}} = \omega_{a} n^{2} r  \sqrt{\dfrac{\hbar \omega_{b}}  {\varepsilon_{0} \varepsilon \mathcal{V}_{b}}},     
\end{equation}
where the interaction mode volume is $ \mathcal{V}_{b} $. Therefore to attain a large vacuum coupling rate $ g_0 $, a large overlap of the electric field distribution and the optical mode of the cavity has to be attained in conjunction with a material with high electro-optic coefficient $r$, high refractive index $n$ and low microwave dielectric constant $ \varepsilon $. The electro-optical interaction is formally equivalent to the optomechanical Hamiltonian, in which a strong pump field of frequency $\omega_p$ enhances $g_0$ in proportion to its amplitude, and the microwave field plays the role of the mechanical motion, see Section III. Consequently, after linearizing the system and in the rotating wave approximation, the interaction term of Eq. \eqref{eq:Hamiltonian} becomes $\hbar g_0 \left( \hat{a}\hat{b}^{\dagger} + \hat{a}^{\dagger}\hat{b} \right)$, which describes a beam splitter interaction (cf. Eq. \eqref{Eq:Heff}). This interaction empties the microwave (optical) mode and upconverts (downconverts) the microwave (optical) photon to an optical (microwave) photon at frequency $\omega_p+\omega_b$ ($\omega_b$).

The electro-optic conversion approach is attractive since it is mechanically and thermally stable (e.g. does not rely on freestanding structures), broadband (for strong electro-optic coefficients), scalable, tunable (e.g., using bias voltages), and (potentially) low noise. Up to now there has been no demonstration of conversion at the quantum level, yet there has been proposals \cite{Tsang_2010, tsang2011, javerzac2016, Soltani_2017, Rueda2019} and some initial demonstrations with strong signals using protocols and approaches that promise coherent quantum conversion \cite{rueda2016, fan2018, witmer2017, Jiang2019}.

One proposal by Javerzac-Galy et al. \cite{javerzac2016} uses integrated superconducting microwave and nonlinear optical microresonators to confine electromagnetic modes to a small volume $\mathcal{V}_{b} \ll \lambda^3$. The integrated device is based on an optical whispering gallery mode microresonator made from a material that features $ \chi^{(2)} $ nonlinearity, such as lithium niobate ($\mathrm{LiNbO}_{3} $) or aluminium-nitride ($ \mathrm{Al}\mathrm{N}$). Their design features a planar optical cavity that is electro-optically coupled to an open superconducting microstrip resonator. Note that the symmetry of the microwave resonator must be broken to ensure that only the positive-component of the phase of the microwave electric field profile couples to the optical microresonator.
Using $\mathrm{LiNbO}_{3}$, in conjuction with high quality factor of optical and microwave cavities of up to $10^6$ and $10^4$ respectively, Javerzac-Galy et al. predict a $g_0/2\pi$ in the tens of kHz with optical pump powers in the mW range.

Concerning fabrication of microresonators from electro-optical materials, one approach is to etch commercially-available crystalline $\mathrm{LiNbO}_{3}$ thin films to allow the combination of a large on-chip density of integrated photonics with the second-order nonlinearity of $\mathrm{LiNbO}_{3}$ \cite{zhang2017}. Optical resonators with quality factors of $\sim 10^6$ have been demonstrated with this approach. We note that the absence of a symmetry center in $\chi^{(2)}$ materials also permits piezoelectricity \cite{thurston1974}. By design, the microring of Ref. \cite{javerzac2016} is embedded in silica ($ \mathrm{SiO}_{2} $) and is thus clamped. Hence, the mechanical degree of freedom is frozen and the piezoelectric contribution to the modulation can be made negligible. 

An approach suggested by Soltani et al. \cite{Soltani_2017} utilizes integrated coupled optical resonators in $\mathrm{SiO}_{2}$-cladded $\mathrm{LiNbO}_{3}$ in conjunction with coplanar microwave resonators. The optical resonator design supports a resonance avoided-crossing doublet with a frequency splitting that matches the resonance frequency of the microwave photon. This proposal features some practical benefits compared to others. Specifically, it allows tuning of the frequency splitting using a DC electro-optic interaction to match the resonance frequency of the microwave cavity, avoiding the necessity of the microwave frequency to match the free-spectral-range of the resonator. This is an approach that significantly increases the dimensions of the resonator and reduces the impact of any undesired conversion or limited that occurs in off-resonant pumping schemes. This scheme offers similar coupling strengths as that of Javerzac-Galy et al. with comparable pump powers and resonators quality factors.  

Notably, a few experiments towards coherent electro-optic transduction at the quantum level have also taken place recently. One by Rueda et al. \cite{rueda2016} demonstrated single-sideband up- or down-conversion of light in a triply resonant whispering gallery mode resonator by addressing modes with asymmetric free spectral range. They showed an electro-optical conversion efficiency of up to 0.1\% photon number conversion for a 10 GHz microwave tone with 0.42 mW of optical pump power and with a bandwidth of 1 MHz. The asymmetry is achieved by avoided crossings between different resonator modes. Despite the large optical quality factor of $10^8$ shown by this scheme, the approach is based on three-dimensional microwave cavities, which limit the optical and microwave mode overlap and the effective electro-optic coupling strength. Nonetheless, larger microwave quality factors of up to $10^5$ suggest $g_0/2\pi$ into the kHz.

The mode overlap issue was addressed in a work by Fan et al. \cite{fan2018}, in which they performed conversion between microwave and optical photons with planar superconducting resonators that are integrated with $ \mathrm{Al}\mathrm{N}$ optical cavities on the same chip. The possibility of the triple-resonance condition with small mode volumes significantly boosted the electro-optic coupling rate which was exemplified by an internal (total) conversion efficiency of 25 (2) \% with a conversion bandwidth of $0.59$ MHz. Furthermore, they observed electromagnetically-induced transparency as a signature of coherent conversion between microwave and optical photons, which was lacking in previous demonstrations, and estimated the number of added microwave noise photons $N_{add}\sim 3$ . Improvements to quality factors with $ \mathrm{Al}\mathrm{N}$ on insulating sapphire substrates suggest efficient conversion.

We also mention that silicon rings, resonators and photonic crystal cavities on $\mathrm{LiNbO}_{3}$ have been demonstrated by Witmer et al. \cite{witmer2017}. Optical quality factors range in the hundreds of thousands, with up to 20\% of the optical mode evanescently coupling to $\mathrm{LiNbO}_{3}$, yet no microwave resonators were featured in their work. 

Finally, there is a non-zero electro-optic contribution to transduction using suspended nonlinear opto-electro-mechanical structures, e.g. those of Ref. \cite{Jiang2019}. 

\section{Other approaches}
The approaches and systems that we discussed above constitute the major part of the effort towards realizing a microwave-to-optical quantum transducer. However, there are some other possible routes that, at least at the moment, are represented less prominently in the field. One of these approaches is for example the magnon based transducer~\cite{Hisatomi_2016}. The underlying idea here is to use collective magnetostatic spin excitations (magnons) as the intermediary mode. Some materials, such as yttrium-iron-garnet (YIG), when put in a homogeneous external magnetic field show distinctive resonance modes for the magnetic(spin) excitations perpendicular to the bias field. For the lowest order resonance mode the excitation is distributed uniformly throughout the material and we can think of it as a large magnetic dipole precessing around the bias field. Due to relatively low damping rate the spin excitation stays coherent for a time long enough to be able to strongly couple to a microwave cavity mode resulting in the hybridized eigenmodes of the coupled system~\cite{Tabuchi_2014,Zhang_2014}. Light coupling is achieved through the pronounced Faraday effect present in YIG crystals. The time modulation of the magnetization caused by the oscillating microwave field creates sidebands to the incidental carrier light allowing for transduction between microwave fields and optical fields in the sidebands. The precession frequency, and hence the magnon resonance frequency, is proportional to the strength of the external magnetic field allowing for an additional tuning knob in the system. Further advantage of the magnon system is its potentially broad conversion bandwidth of the order of few MHz. Maximum conversion efficiency achieved in an experiment was around $10^{-10}$ and was mainly limited by the weak light-magnon coupling~\cite{Hisatomi_2016}, which could be enhanced by placing optical cavity around the YIG crystal.

Instead of using the direct conversion of microwave photons to optical ones and vice versa, one can transfer the quantum state of one photon to another using entanglement and quantum teleportation. A necessary condition for that is an entangled state between the optical and the microwave photons. In almost every of the previously discussed systems by adjusting the system parameters one can change the effective Hamiltonian from a beam-splitter like Hamiltonian \eqref{eq:Hamiltonian} to a two-mode squeezing like Hamiltonian
\begin{align}
    H_{eff} = \hbar\Omega\tilde{g}_{eff}\hat{a}^\dagger \hat{b}^\dagger + h.c. \label{eq:TMSHamiltonian}.
\end{align}
As already pointed out in Sec. \ref{seq:OptomechanicalTheory} free evolution under such Hamiltonian creates a so called two-mode squeezed state that can be used for state transfer using continuous variable teleportation~\cite{Barzanjeh_2012}. Operated in pulsed regime the same Hamiltonian could also be used to create discrete variable entanglement using post-selection. For example, Ref.~\cite{Zhong_2019} discusses realization of the time-bin entanglement between microwave and optical photons using underlying Hamiltonian.

Coherent transfer of phase information between optical and microwave fields was proposed in Ref.~\cite{Lekavicius_2017}, where a single NV center was used as a mediator. The relative phase between two microwave fields was encoded into a coherent superposition between the ground spin triplet states $|+1\rangle$ and $|-1\rangle$ and read out using two light fields with opposite circular polarization and Raman transition.

Another approach is based on the emitters that have permanent dipole moments, such as organic dye molecules or quantum dots, and are embedded in a phononic waveguide to enhance the light-matter coupling~\cite{Das_2017, Elfving_2018}. If a two level system that has dipole allowed transition is placed in the proximity of such an emitter, the electric field associated with this dipole transition can interact with the permanent dipole moment of the emitter leading to a state dependent Stark shift. If this shift in the transition frequency is larger than the transition linewidth this configuration allows one to directly entangle the two level system, e.g. superconducting qubit, with a scattered optical photon, whose frequency will depend on the state of the two level system, without the detour via microwave photons. One can enhance the emission probability of the entangled optical photon by placing two of such emitters in the vicinity of each other. The presence of permanent dipole moments and the close distance result in a strong dipole-dipole interaction between the two emitters leading to hybridized eigenstates that interact more strongly with the two level system. 

The use of variations in graphene optical conductivity in response to externally applied fields has been suggested by Qasymeh and Eleuch \cite{qasymeh2019}.



\section{Discussion and Outlook}

We have reviewed different approaches to quantum transduction. The field is still in an early stage, but there is a lot of activity and a lot of progress. 

How close are we to having useful transducers and what are the metrics that could be used to assess their performance? The answer may depend on what application one has in mind. 

For example, if the goal is to use the transducer in the context of single photon detection either by converting a microwave photon to the optical domain, then detecting the optical photon, or conversely using microwave photons and superconducting qubits for quantum non-demolition measurement of optical photons, then the overall transduction efficiency will be one of the key features.  

On the other hand, if the goal is to use optical photons to entangle distant superconducting qubits or microwave cavities (e.g. for distributed quantum computing or quantum repeaters), then it is likely to be more important that these photons are indistinguishable, so that they remain suitable for single-photon or two-photon interference to ensure high fidelity entanglement generation, whereas the conversion efficiency would be less critical, although it shouldn't be too low either, if the goal is to achieve reasonable rates.

For both examples it would be important that there are very little added noise photons requiring high signal-to-noise ratio of the transducer. 

\begin{figure}[t]
 	\includegraphics[width=1.05\columnwidth]{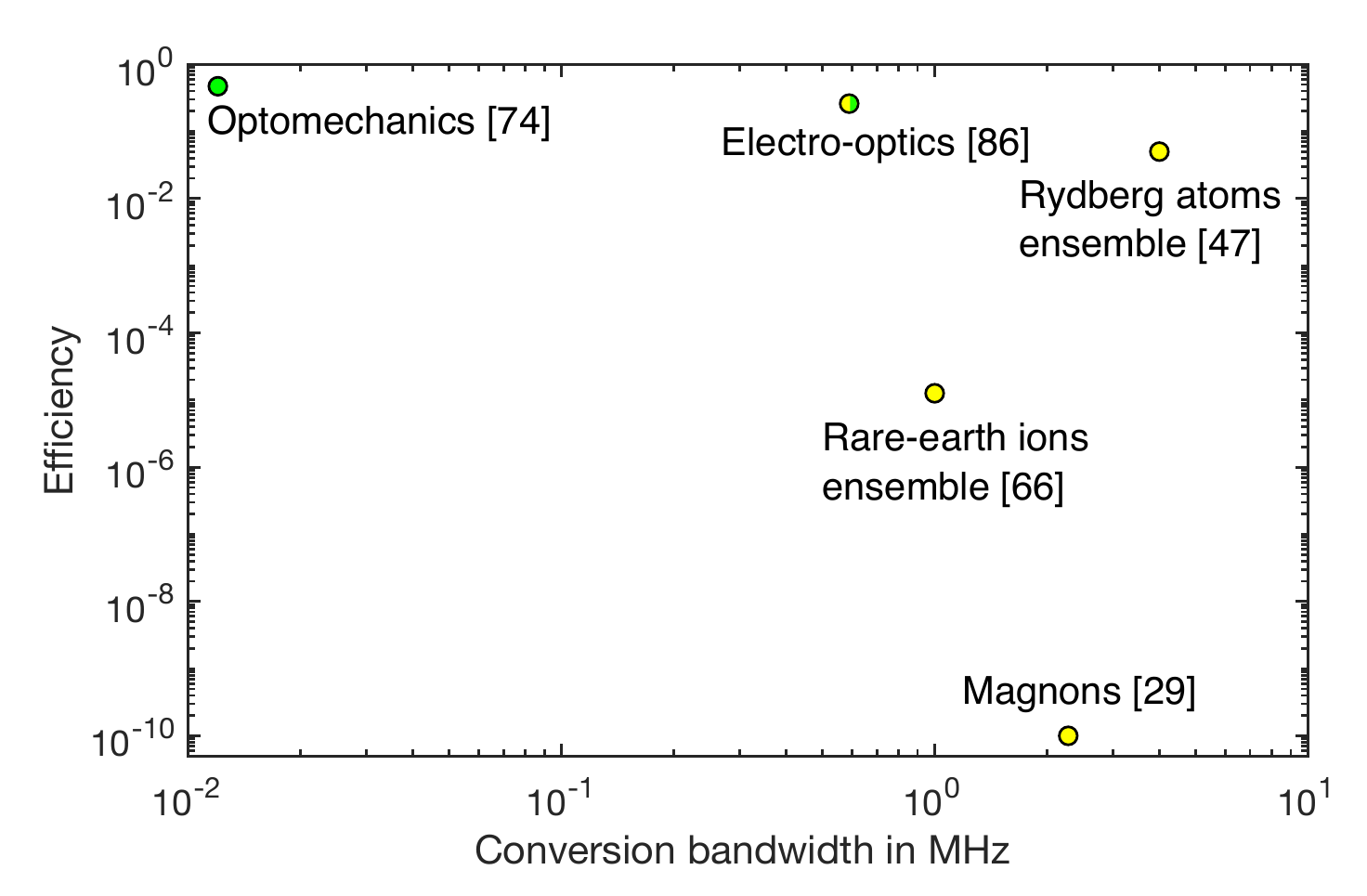}
 	\caption{Maximally achieved conversion efficiency and conversion bandwidth for the different transduction approaches. The efficiency value for the electro-optics is the internal conversion efficiency, in the case of Rydberg atoms ensemble the efficiency value is given for a conversion of a free space microwave signal. Green coloring indicates proposals that are almost in the quantum transduction regime and where added noise was measured. Green-yellow coloring indicates proposals where the number of added noise photons was estimated based on the model used to describe the system.}
 	\label{Fig:EffBandwidth}
\end{figure}

Conversion bandwidth is another figure of merit that might be of practical importance, in particular for applications where the transducer is likely to be the rate-limiting element. Since high bandwidth is a necessary requirement for frequency and time multiplexing that would allow to increase the operation rate. However, bandwidth may be less critical for other applications, e.g. in the context of long-distance quantum communication (quantum repeaters), where rates are often limited by other factors, such as communication times due to the finite speed of light.
Moreover, conversion bandwidth is limited by the GHz resonance frequencies of microwave qubits.
In Fig. \ref{Fig:EffBandwidth} we plot the maximally achieved transduction efficiencies for the different approaches together with the corresponding conversion bandwidth. As of now only optomechanically based experiments are able to perform the microwave-to-optical transduction at the few photon level, all the other reported conversion efficiencies were measured for classical signals.

Despite being implemented in different physical platforms and despite the fact that some of the approaches are more advanced than others, many of them face similar challenges that should be resolved moving towards \textit{quantum} transducers. For example all of the approaches rely on a strong light-matter interaction that allows for a coherent transfer of optical excitation to some kind of matter excitation (spin wave or resonator oscillation), just as is the case in optical quantum memories. In contrast to the quantum memories however, transduction doesn't require these matter excitation to be long-lived. On the other hand, transduction also requires relatively strong coupling to microwaves, at least an order of magnitude larger than the relevant coherence times, in order to faithfully transduce quantum states of the photons. This implies that microwave cavities with high quality factor should be used. Microwave cavities with the highest Q-factors are obtained using superconducting materials \cite{Romanenko_2018}.

Another common aspect in all of the discussed approaches is the use of strong optical fields to bridge the energy gap between optical and microwave photons and the necessity to operate at mK temperatures to suppress the number of thermal microwave photons. Both of these requirements are to some extent at odds to each other, since strong optical fields often cause absorption-induced heating or nonlinearity, or noise due to spectator atoms. Scattered laser light can also destroy the superconductivity of the cavity material resulting in the decrease of the cavity's Q-factor. One possible route to address this issue is to use small mode volume optical cavities to reduce laser power requirements. However, one has to ensure that strong microwave coupling is still achievable with such smaller systems. 

Since some of the approaches are more advanced than others the next steps to proceed will depend on the underlying system. For example, a next important step for the atomic clouds would be demonstration of microwave-to-optical conversion using microwave cavities or waveguides instead of classical free space microwave fields. In the case of rare-earth ions doped crystals the next natural step is to show the performance at mK temperatures, and for the currently most advanced candidate -- optomechanically based transduction -- the next steps include reducing the number of noise photons by optimizing feedback control mechanism and by using mechanical oscillators with higher Q-factors.

To summarize, the quantum transduction of microwave and optical fields is currently a very active area of research. There has been a lot of progress in a relatively short amount of time, with the best systems operating already at the few-photon level with relatively high efficiency. However, this challenging endeavour is far from being completed, with a lot of interesting physics still lying ahead of us.

\subsection*{Acknowledgments}
During the writing of this article we became aware of another review of quantum transduction with somewhat different emphasis \cite{Lambert_2019}. 

We would like to thank the participants of the transduction workshop at Caltech in September 2018 for helpful and stimulating discussions. We particularly thank John Bartholomew, Andrei Faraon, Johannes Fink, Jeff Holzgrafe, Linbo Shao, Marko Lon\v{c}ar, Daniel Oblak, and Oskar Painter.

N.L. and N.S. acknowledge support from the Alliance for Quantum Technologies' (AQT) Intelligent Quantum Networks and Technologies  (INQNET) research program and by DOE/HEP QuantISED program grant, QCCFP (Quantum Communication Channels for Fundamental Physics), award number DE-SC0019219.
N.S. further acknowledges support by the Natural Sciences and Engineering Research Council of Canada (NSERC).
S.B. acknowledges support from the Marie Sk\l{}odowska Curie fellowship number 707438 (MSC-IF SUPEREOM).
J.P.C. acknowledges support from the Caltech PMA prize postdoctoral fellowship.
M.S. acknowledges support from the ARL-CDQI and the National Science Foundation. 
C.S. acknowledges NSERC, Quantum Alberta, and the Alberta Major Innovation Fund.  

\bibliographystyle{apsrev4-1}
\bibliography{Mw2OptRev}

\end{document}